\documentclass[aps,nofootinbib,12pt]{revtex4}
\usepackage{amsmath}
\usepackage{amsfonts}
\usepackage{amssymb}
\usepackage{color}
\usepackage{graphicx}
\usepackage{mathrsfs}
\usepackage{epstopdf}
\usepackage{latexsym}

\def\bc{\begin{center}}
\def\nno{\nonumber}
\def\ec{\end{center}}
\def\be{\begin{eqnarray}}
\def\ee{\end{eqnarray}}

\definecolor{dyellow}{rgb}{1.,0.8,.0}
\definecolor{myblue}{rgb}{.1,.1,.7}
\definecolor{dcyan}{rgb}{.0,.6,.6}
\definecolor{dmagenta}{rgb}{0.6,0.0,0.6}
\definecolor{brown}{rgb}{0.6,0.2,0.}
\definecolor{darkblue}{rgb}{.0,.0,0.5}
\definecolor{darkred}{rgb}{0.75,0.0,0.0}
\definecolor{orange}{rgb}{1.,.6,.0}
\definecolor{dorange}{rgb}{0.8,.4,.0}
\definecolor{darkgreen}{rgb}{0.0,0.6,0.0}
\definecolor{purple}{rgb}{.4,.0,.4}
\definecolor{lightgrey}{rgb}{0.7, 0.7, 0.7}
\definecolor{grey}{rgb}{0.4, 0.4, 0.4}

\def\al{\alpha}

\def\si{\sigma}
\def\om{\omega}

\def\La{\Lambda}

\begin{document}

\title{Holographic p-wave superconductors from Gauss-Bonnet gravity} \vskip 2cm \vskip 2cm
\author{Rong-Gen Cai}\email{cairg@itp.ac.cn}
\author{Zhang-Yu Nie}\email{niezy@itp.ac.cn}
\author{Hai-Qing Zhang}\email{hqzhang@itp.ac.cn}
\address{Key Laboratory of Frontiers in Theoretical Physics,
Institute of Theoretical Physics, Chinese Academy of Sciences,
   P.O. Box 2735, Beijing 100190, China}


\begin{abstract} \vspace{5mm}
 We study the holographic p-wave superconductors in a five-dimensional Gauss-Bonnet gravity with an SU(2)
 Yang-Mills gauge field. In the probe approximation, we find that when the Gauss-Bonnet coefficient
grows, the condensation of the vector field becomes harder, both the
perpendicular and parallel components, with respect to the direction
of the condensation, of the anisotropic conductivity decrease. We
also study the mass of the quasi-particle excitations, the gap
frequency and the DC conductivities of the p-wave superconductor.
All of them depend on the Gauss-Bonnet coefficient. In addition we
observe a strange behavior for the condensation and the relation
between the gap frequency and the mass of quasi-particles when the
Gauss-Bonnet coefficient is larger than $9/100$, which is the upper
bound for the Gauss-Bonnet coefficient from the causality of the
dual field theory.

\end{abstract}
\maketitle
\section{Introduction}
The AdS/CFT
correspondence~\cite{Maldacena:1997re,Gubser:1998bc,Witten:1998qj,Aharony:1999ti}
provides a theoretical method to understand strongly coupled field
theories. It has been applied to calculate transport coefficients,
such as shear viscosity, of strongly coupled systems and some
universal properties of dual strongly coupled field theories have
been found in the hydrodynamical
limit~\cite{Policastro:2001yc,Kovtun:2003wp,Buchel:2003tz,Kovtun:2004de}.
Recently, it has been proposed that the AdS/CFT correspondence also
can  be used to describe superconductor phase
transition~\cite{Gubser:2008px,Hartnoll:2008vx}. Since the high
$T_c$ superconductors are shown to be in the strong coupling regime,
one expects that the holographic method could give some insights
into the pairing mechanism in the high $T_c$ superconductors.

There have been lots of work studying various holographic
superconductors
\cite{Hartnoll:2008kx,Horowitz:2008bn,Nakano:2008xc,Basu:2008st,
Herzog:2008he,Roberts:2008ns,Ammon:2008fc,Basu:2008bh,Amado:2009ts,
Koutsoumbas:2009pa,Maeda:2009wv,Basu:2009vv,Sonner:2009fk,Cai:2009hn,
Gubser:2009qm,Gauntlett:2009dn,Horowitz:2009ij,
Konoplya:2009hv,Brynjolfsson:2009ct,Sin:2009wi,Hartnoll:2009sz,
Herzog:2009xv,Albash:2008eh,Wen:2008pb,Gubser:2008zu}, in which some
effects such as scalar field mass, external magnetic field, and back
reaction {\it etc.} have been discussed. Among those works, some
universality is discovered. For instance, the ratio of gap frequency
over critical temperature $\omega_g/T_c$ is found to be always near
the value $8$~\cite{Horowitz:2008bn}, which is more than twice the
weakly coupled BCS theory value $3.5$. The reason for this bigger
value might be that the holographic superconductor is strongly
coupled.

In the holographic study of shear viscosity, the universal bound on
the ratio of the shear viscosity over entropy desnity $\eta/s\geq
1/4\pi$ is found to be violated in theories dual to gravity systems
with some higher curvature
corrections~\cite{Brigante:2007nu,Brigante:2008gz,KP,
Brustein:2008cg,Iqbal:2008by,Cai:2008ph,Cai:2009zv,Buchel:2010wf}.
This promotes the study of holographic superconductors under higher
curvature
corrections~\cite{Gregory:2009fj,Pan:2009xa,Ge:2010aa,BH,PW,New}.
Another motivation to consider the higher curvature effect on the
holographic superconductors is due to the Mermin-Wagner theorem or
Coleman theorem, which states that in quantum field theory,
continuous symmetries cannot be spontaneously broken at finite
temperature in systems with sufficiently short-range interactions in
spatial dimensions $d \le 2$. However, one indeed observes the
superconducting phase transition in the gravity dual of
four-dimensional AdS black hole
backgrounds~\cite{Gubser:2008px,Hartnoll:2008vx}. This might be
caused by the suppression of the large fluctuations in the large N
limit, which is supposed to be one of conditions for the validness
of the AdS/CFT correspondence. In \cite{Gregory:2009fj,Pan:2009xa}
the effect of the Gauss-Bonnet term on the  holographic s-wave
superconductors has been investigated. It is found that a larger
Gauss-Bonnet term makes condensation harder, the universality of
$\omega_g/T_c\approx8$ is violated and the ratio of gap frequency
over critical temperature depends on the Gauss-Bonnet coefficient.

Besides the intensively studied holographic s-wave superconductors,
there also exist holographic p-wave superconductor
models~\cite{Gubser:2008wv}. In the p-wave case, there is a special
direction which breaks the rotational symmetry. In
Ref.~\cite{Gubser:2008wv}, the authors studied the p-wave
superconductors  and observed  an anisotropic conductivity. Their
results fit well with the Drude model in the low frequency limit.
Further the authors of \cite{Ammon:2008fc} studied the back reaction
effect of matter field on the p-wave superconductor and found that
when the ratio of the five-dimensional gravitational constant to the
Yang-Mills coupling is beyond a critical value ($\approx 0.365)$,
the phase transition becomes first order.

In this paper, we are interested in the effect of the Gauss-Bonnet
term on the $4$-dimensional p-wave superconductors, which are dual
to $5$-dimensional Gauss-Bonnet-AdS black holes in the bulk. In the
probe approximation, the bulk spacetime is a $5$-dimensional
Gauss-Bonnet-AdS black hole with a Ricci flat
horizon~\cite{Cai:2001dz}. We find that the condensation increases
as the Gauss-Bonnet coefficient grows, which means a positive
Gauss-Bonnet term makes the condensation harder as the case of
s-wave superconductors.  There are two different kinds of
conductivity due to the anisotropic condensation in p-wave
superconductors. The conductivity perpendicular to the direction of
the condensation behaves like a s-wave one. On the other hand, the
conductivity parallel to the direction of the condensation behaves
much different. In the low frequency regime, this conductivity can
be well explained by Drude model as in Ref.~\cite{Gubser:2008wv}. In
addition, we see that the Gauss-Bonnet term will not change the
order of the phase transition. Namely the phase transition is still
second order and some critical exponents still take their mean-field
theory values.

This paper is organized as follows. In Sec.~\ref{sect:background},
we give out the basic setup and study the superconductor phase
transition in the probe limit with various values of the
Gauss-Bonnet coefficient. In Sec.~\ref{sect:conduc}, we calculate
the anisotropic frequency dependent conductivity. We conclude our
paper in Sec.~\ref{sect:conclu}.

\section{Holographic p-wave superconductor}
\label{sect:background}

 We consider the bulk theory of the Einstein-Gauss-Bonnet gravity
with an SU(2) Yang-Mills field in a $5$-dimensional  space-time. The
action is
\begin{equation}\label{action}
S=\int d^5 x \sqrt{-g} \Big[\frac{1}{2
\kappa_5^2}\Big(R+\frac{12}{L^2}+\frac{\alpha}{{2}} (R^2-4R^{\mu\nu}
R_{\mu\nu}+R^{\mu\nu\rho\sigma}R_{\mu\nu\rho\sigma})\Big){-}\frac{1}{4
\hat{g}^2}\Big(F^a_{\mu\nu}F^{a\mu\nu}\Big)\Big],
\end{equation}
where $\kappa_5$ is the five dimensional gravitational constant,
$\hat{g}$ is the Yang-Mills coupling constant and $L$ the radius of
the AdS spacetime. $F^a_{\mu\nu}=\partial_\mu A^a_\nu-\partial_\nu
A^a_\mu + \epsilon^{abc}A^b_\mu A^c_\nu$ is the Yang-Mills field
strength, $\epsilon^{abc}$ is the totally antisymmetric tensor with
$\epsilon^{123}=+1$. The quadratic curvature term is the
Gauss-Bonnet term with $\alpha$ the Gauss-Bonnet coefficient.

We are interested in the asymptotic AdS solution to this gravity
system. In general, we should solve the Yang-Mills equations as well
as gravitational field equations to search for a required solution.
To solve this problem with the Gauss-Bonnet term will be difficult.
However, we can get some qualitative features in the so-called probe
limit, where the back reaction of matter fields (Yang-Mills field)
on the metric can be neglected. This approximation can be justified.
Indeed, we can see from the action that in the limit
$\kappa_5^2/\hat{g}^2\ll 1$, the back reaction of the Yang-Mills
field on the metric can be neglected safely.

In the probe limit, the background metric is a $5$-dimensional
Gauss-Bonnet-AdS black hole with a Ricci flat
horizon~\cite{Cai:2001dz}. The metric is described by
\begin{eqnarray}\label{metric}
ds^2
&=&-f(r)dt^2+\frac{1}{f(r)}dr^2+\frac{r^2}{L^2}(dx^2+dy^2+dz^2),
\\
f(r) &=&\frac{r^2}{2\alpha}\bigg(1-\sqrt{1-\frac{4\alpha}{L^2}(
1-\frac{mL^2}{r^4})}\bigg), \nonumber
\end{eqnarray}
where $m$ is the mass of the black hole. The horizon is located at
$r=r_h=\sqrt[4]{mL^2}$, and the temperature of the black hole is
\begin{equation}\label{temperatur}
T=\frac{r_h}{\pi L^2}.
\end{equation}
Here we should notice that in the asymptotic region with $r \to
\infty$,
\begin{equation}
f(r)\sim
\frac{r^2}{2\alpha}\Bigg[1-\sqrt{1-\frac{4\alpha}{L^2}}\Bigg].
\end{equation}
One can define an effective radius $L_{\text{eff}}$ of the AdS
spacetime by
\begin{equation}\label{L2}
L_{\text{eff}}^2\equiv\frac{2\alpha}{1-\sqrt{1-\frac{4\alpha}{L^2}}}.
\end{equation}
We can see from this equation that one has to have $\alpha\leq
L^2/4$ in order to have a well-defined vacuum for the gravity
theory. The upper bound $\alpha =L^2/4$ is called Chern-Simons
limit. In the AdS/CFT correspondence, this asymptotic AdS spacetime
is dual to a conformal field theory living on the boundary
$r\rightarrow\infty$. The temperature of the black hole is just the
one of the dual field theory. If we further consider the causality
constraint from the boundary CFT, there is an additional constraint
on the Gauss-Bonnet coefficient with $-7L^2/36\leq \alpha \leq
9L^2/100$ in five
dimensions~\cite{Brigante:2007nu,Brigante:2008gz,Buchel:2009tt,
Hofman:2009ug,deBoer:2009pn,Camanho:2009vw,Buchel:2009sk}.

The probe SU(2) Yang-Mills field $A^a_{\mu}$ is dual to some current
operator $J_{\mu}^a$ in the $4$-dimensional boundary theory. In
order to realize a holographic p-wave superconductor, following
Ref.~\cite{Gubser:2008wv,Manvelyan:2008sv} we adopt the ansatz
\begin{equation}\label{ansatz}
A=\phi(r)\tau^3dt+\psi(r)\tau^1dx
\end{equation}
for the Yang-Mills gauge field. Here $\tau^i$s are the three SU(2)
generators with commutation relation
$[\tau^i,\tau^j]=\epsilon^{ijk}\tau^k$. In this ansatz one can
regard the U(1) subalgebra generated by $\tau^3$ as the gauge group
of electromagnetism, and then the condensation of $\psi(r)$ will
break the U(1) symmetry and lead to the superconductor phase
transition. Because $\psi(r)$ is dual to the $J_x^1$ operator on the
boundary, choosing x-axis as a special direction, the condensation
of $\psi(r)$ breaks the rotational symmetry and leads to a phase
transition, which can be interpreted as a p-wave superconducting
phase transition on the boundary.

The Yang-Mills equations with the above ansatz (\ref{ansatz}) are
\begin{equation}\label{ym}
\left\{ \begin{aligned} \phi''+\frac{3 }{r}\phi '-\frac{L^2
\psi^2}{r^2 f}
\phi=0,\\
\psi ''+(\frac{1}{r}+\frac{f' }{f})\psi '+\frac{\phi ^2
}{f^2}\psi=0.
\end{aligned}\right.
\end{equation}
We'll solve the Yang-Mills equations on the black hole background
(\ref{metric}) and study the solutions with non-zero $\psi(r)$ which
is related to the p-wave superconducting phase.

In order to solve the equations (\ref{ym}), we need the boundary
conditions for the fields $\psi(r)$ and $\phi(r)$. On the black hole
horizon, it is required that $\phi(r_h)=0$ for the U(1) gauge field
to have a finite norm, and $\psi(r_h)$ should be finite. Therefore,
the boundary conditions of $\psi$ and $\phi$ on the horizon are:
 \be \psi&=&\psi_H^{(0)}+\psi_H^{(2)}(1-\frac{r_h}{r})^2+\cdots,\\
 \phi&=&\phi_H^{(1)}(1-\frac{r_h}{r})+\cdots.\ee

 On the boundary of the bulk, we have
\begin{eqnarray}
\phi(r)&\rightarrow& \mu + \rho/r^2 \\
\psi(r)&\rightarrow& \psi^{(0)} + \psi^{(2)}/r^2.
\end{eqnarray}
$\mu$ and $\rho$ are dual to the chemical potential and charge
density of the boundary CFT , $\psi^{(0)}$ and $\psi^{(2)}$ are dual
to the source and expectation value of the boundary operator $J_x^1$
respectively. We always set the source $\psi^{(0)}$ to  zero, as we
want to have a normalizable solution.

The trivial solution to the above Yang-Mills equations is a charged
black hole solution with $\psi(r)=0$, which is just the
non-superconducting phase in the boundary theory. We will try to
find non-trivial solutions,  describing the p-wave superconducting
phase, with $\psi(r)\neq0$. We use a shooting method in which we
solve the Yang-Mills equations numerically from the horizon to
boundary, and pick the suitable one obeying the boundary conditions
at $r\rightarrow\infty$.  In the numerical calculation, we set $L=1$
and define a new  variable $z=r_h/r$. Rescaling $r$, $\phi(r)$ and
$\psi(r)$, one then can simply set $r_h=1$.

\begin{figure}
\includegraphics[width=13cm] {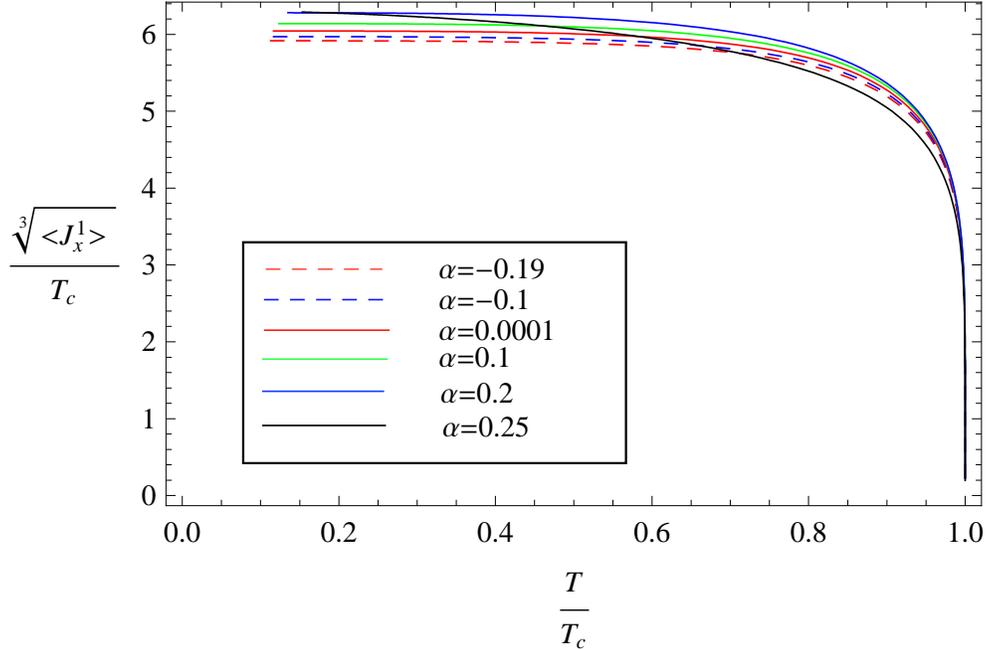}
\caption{\label{Condensation} The condensation of $J_x^1$ as a
function of temperature. The six lines correspond to different
Gauss-Bonnet coefficient, respectively: The dashed red line to
$\alpha=-0.19$, the dashed blue line to $\alpha=-0.1$, the red line
to $\alpha=0.0001$, the green line to $\alpha=0.1$, the blue line to
$\alpha=0.2$, and the black line to $\alpha=0.25$ which saturates
the Chern-Simons limit. }
\end{figure}

In Figure~\ref{Condensation}, we plot the condensation of $J_x^1$ as
a function of temperature with various values of the  Gauss-Bonnet
coefficient $\alpha$. Note that the boundary operator $J_x^1$ is a
component of a vector operator.  Unlike the scalar field in the
holographic s-wave superconductor, one can find in the AdS/CFT
dictionary~\cite{Aharony:1999ti} that the conformal dimension of
$J_x^1$ is $\lambda=3$, the same as the charge density $\rho$. In
the figure, we therefore plot the data of $J_x^1$ as
$\sqrt[3]{J_x^1}/T_c$, as this is the right dimensionless quantity.
We can see from the figure that the condensation value increases
with the increase of the Gauss-Bonnet coefficient. This is the same
as the s-wave case~\cite{Gregory:2009fj,Pan:2009xa}. Note that the
curve for the case $\alpha=0.25$ intersects with others. We have
some to say later on this behavior.
\begin{figure}
\includegraphics[width=13cm] {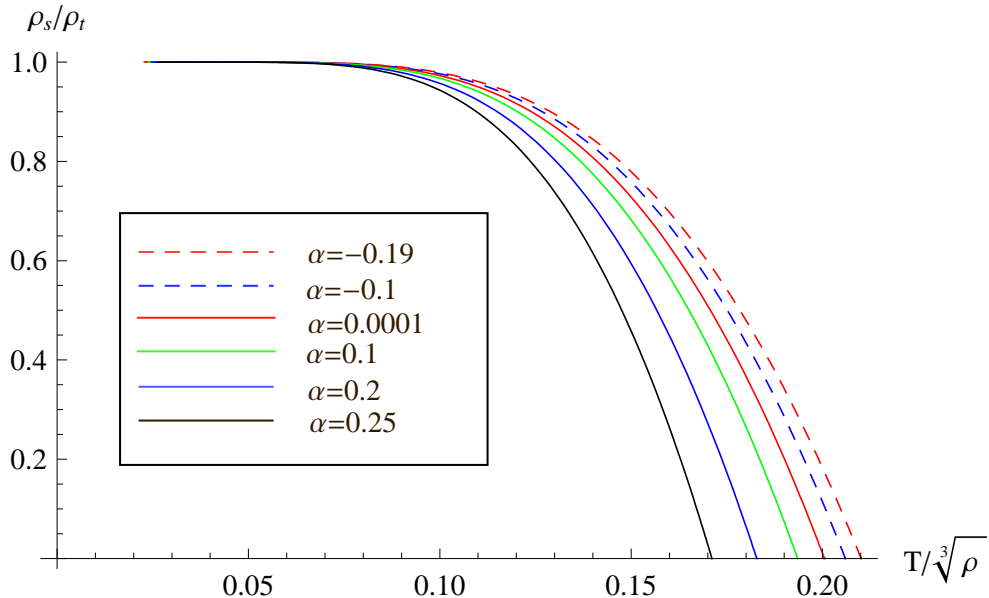}
\caption{\label{density} The ratio of the superconducting charge
density to the total charge density versus temperature. The
intersecting points of the curves with the horizontal axis represent
the critical temperature $T_c$ when the superconducting phase
occurs. $T_c$ decreases when $\al$ increases. }
\end{figure}

 In  mean-field theory, the condensation of operator is
proportional to $\sqrt{(1-T/T_c)}$ when $T\rightarrow T_c$. We can
see from the figure \ref{Condensation} that the p-wave condensations
have a similar behavior like that in mean-field theory when
$T\rightarrow T_c$, we fit the data as follows:
\begin{eqnarray}
\al=-0.19, &&  {\langle J^1_x\rangle}=476.2698~ T_c ^3(1 -
T/T_c)^{1/2}, \nonumber \\
\al=-0.1, && \langle J^1_x\rangle=487.0347 ~T_c^3 (1 -
T/T_c)^{1/2}, \nonumber \\
\al=0.0001, && \langle J^1_x\rangle=499.0358 ~T_c^3(1 -
T/T_c)^{1/2}, \nonumber \\
\al=0.1, && \langle J^1_x\rangle=509.7952 ~T_c^3 (1 - T/T_c)^{1/2}, \nonumber \\
\al=0.2, && \langle J^1_x\rangle=519.5827 ~T_c^3(1 - T/T_c)^{1/2}, \nonumber \\
\al=0.25, && \langle J^1_x\rangle=422.3983~ T_c^3(1 - T/T_c)^{1/2}.
\end{eqnarray}
This indicates that the Gauss-Bonnet term does not change the
critical exponent of the condensation.

In the holographic p-wave superconductor, the normal charge density
$\rho_n$ is $\rho_n=\phi_H^{(1)}$, and the total charge density is
$\rho_t=2\rho$, the factor $2$ arises due to the scaling behavior of
$\phi$ on the infinite boundary. The superconducting charge density
is defined as $\rho_s=\rho_t-\rho_n$. We plot the ratio
$\rho_s/\rho_t$ in Figure~\ref{density}. The points where the curves
intersect with the horizontal axis represent the critical
temperatures for different $\al$ when the superconducting phase
occurs. We can see from the figure that the critical temperature
decreases with the increase of the Gauss-Bonnet coefficient. In
fact,   \be
  T_c=0.2101\sqrt[3]{\rho} &&\text{when}\quad \alpha=-0.19,\nno\\
  T_c=0.2060\sqrt[3]{\rho} &&\text{when}\quad \alpha=-0.1,\nno\\
  T_c=0.2005\sqrt[3]{\rho} &&\text{when}\quad \alpha=0.0001,\nno\\
 T_c=0.1935\sqrt[3]{\rho} &&\text{when}\quad \alpha=0.1,\nno\\
 T_c=0.1828\sqrt[3]{\rho} &&\text{when}\quad \alpha=0.2,\nno\\
 T_c=0.1711\sqrt[3]{\rho} &&\text{when}\quad \alpha=0.25. \ee
So we conclude that a  larger Gauss-Bonnet term makes the
condensation harder to form. This result is qualitatively the same
as the Gauss-Bonnet effect on the holographic s-wave
superconductors~\cite{Gregory:2009fj,Pan:2009xa,Ge:2010aa}.

In the next section, we will calculate the frequency dependent
conductivity to see the influence of the Gauss-Bonnet term more
clearly.

\section{Electric Conductivity}
\label{sect:conduc}

In order to see the electric conductivity of the system, we can add
an electromagnetic perturbation into the system. For the Yang-Mills
case, the perturbation of the electric field will mix other
components in the linearized equation \cite{Gubser:2008wv}. The
perturbation is $A\rightarrow A+\delta A$, where
 \begin{equation}\label{infesimal} \delta A= e^{-i\omega t}[(A^1_t(r)\tau^1+A^2_t(r)\tau^2)dt
 +A^3_x(r)\tau^3dx+A^3_{\text{y}}(r)\tau^3dy].
 \end{equation}
Note that there still exists a SO(2) symmetry in the plane $y-z$ in
the system. The electric conductivity $\sigma_{\text{zz}}$ is
completely the same as $\sigma_{\text{yy}}$. Therefore we have
neglected the perturbation along the direction $z$ for simplicity.
In the following subsections, we separately calculate the components
$\sigma_{\text{yy}}$ and $\sigma_{\text{xx}}$ to see the difference
between them, and study the effect of the Gauss-Bonnet term on the
electric conductivities.

\subsection{$\sigma_{\text{yy}}$}

 The linearized equation of motion for $A_{\text{y}}^3$
decouples from other components of the Yang-Mills field. The
equation of motion is
\begin{equation}\label{yyEom}
(\frac{\omega ^2 }{f^2}-\frac{L^2 \psi ^2}{r^2
f})A^3_{\text{y}}+(\frac{1}{r}+\frac{
f'}{f}){A^3_{\text{y}}}'+{A^3_{\text{y}}}''=0.
\end{equation}
The equation (\ref{yyEom}) is very similar to  corresponding
equation in the holographic s-wave superconductors
\cite{Gubser:2008px,Hartnoll:2008vx,Gubser:2008wv}. Therefore, the
calculation of $\sigma_{\text{yy}}$ is the same as that in the
s-wave case.  But we still show its details here  in order to make a
contrast to $\sigma_{\text{xx}}$.

To calculate the conductivity, we impose the in-falling wave
condition to $A_{\text{y}}^3$ on the horizon. Then the current
Green's function with zero spatial momentum $G^R(\omega,\vec{0})$
can be evaluated using AdS/CFT correspondence as
\cite{son,Horowitz:2009ij}
\begin{equation}
G^R(\omega,\vec{0})=-\lim\limits_{r\to \infty} r f  A_{\text{y}}^3
A_{\text{y}}^{3}{'}.
\end{equation}
Here, $A^3_y(r)$ is normalized to be $A^3_y(r\rightarrow\infty)=1$.
Near the boundary of the AdS bulk, the expansion of $A^3_y$ is
\begin{equation}
A^3_{\text{y}}=A_{\text{y}}^{3(0)}+\frac{A_{\text{y}}^{3(2)}}{r^2}+\frac{A_{\text{y}}^{3(0)}\omega^2
L_{\text{eff}}^2}{2} \frac{\text{log} \Lambda r}{r^2}.
\end{equation}
where, $L_{\text{eff}}^2$ is the one in the formula (\ref{L2}), and
$\Lambda$ is an arbitrary constant. Thus the retarded Green's
function is
\begin{equation}
G^R(\omega,\vec{0})=2\frac{A_{\text{y}}^{3(2)}}{A_{\text{y}}^{3(0)}
L_{\text{eff}}^2} +\omega^2 L_{\text{eff}}^2 (\text{log} \Lambda
r-\frac{1}{2}).
\end{equation}
The logarithmic divergence can be removed with a boundary
counterterm in the gravity action \cite{Taylor:2000xw}. This
procedure is related to renormalization of the UV divergence in the
boundary theory. After adding a counterterm to cancel the
logarithmic divergence, we have the retarded Green's function
\begin{equation}
G^R(\omega,\vec{0})=2\frac{A_{\text{y}}^{3(2)}}{A_{\text{y}}^{3(0)}
L_{\text{eff}}^2} -\frac{1}{2}\omega^2 L_{\text{eff}}^2 ,
\end{equation}
and the conductivity is
\begin{equation}\label{yyconductivity}
\sigma_{\text{yy}}=\frac{1}{i\omega}G^R(\omega,\vec{0})=\frac{-2 i
A_{\text{y}}^{3(2)}}{A_{\text{y}}^{3(0)} \omega L_{\text{eff}}^2}
+\frac{i}{2}\omega L_{\text{eff}}^2.
\end{equation}

\begin{figure}
\includegraphics[width=8cm] {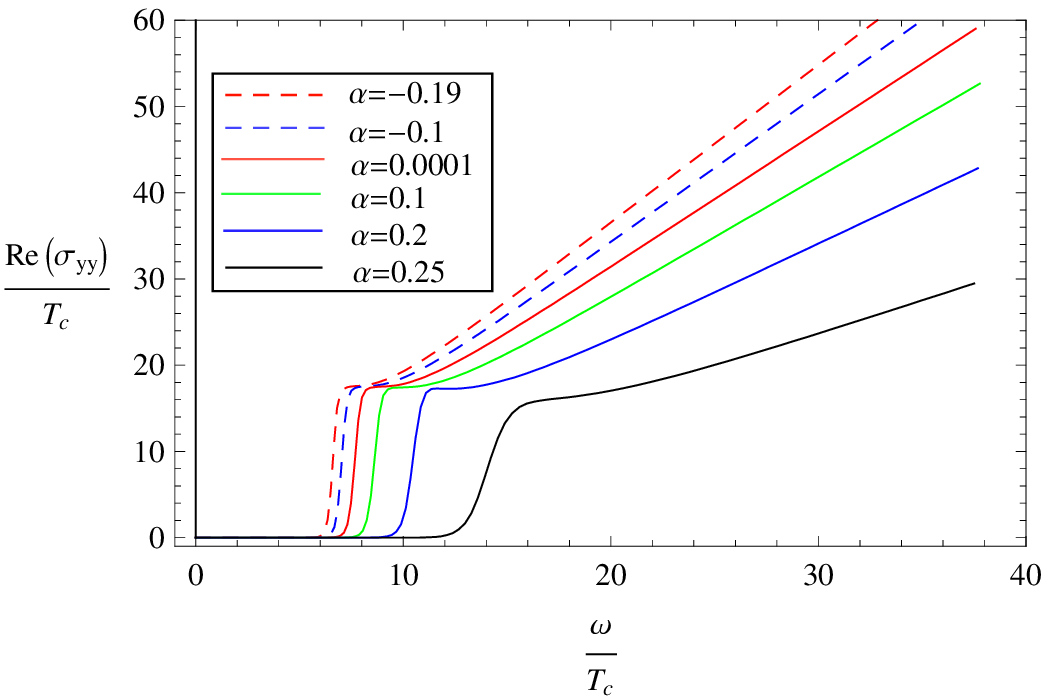}
\includegraphics[width=8cm] {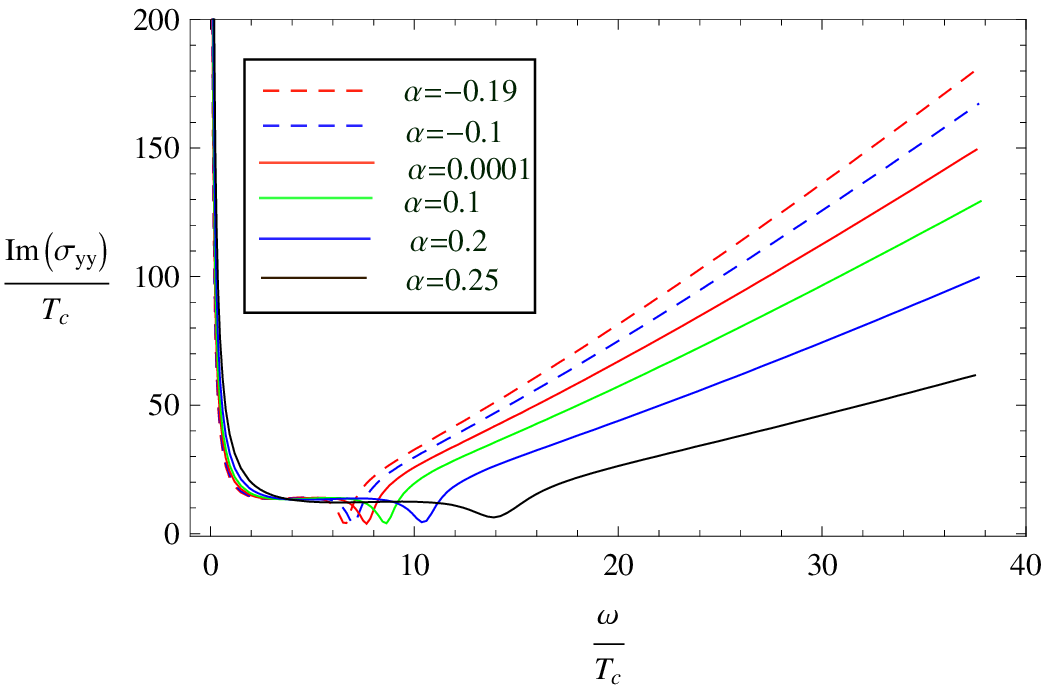}
\caption{\label{Condyy} The conductivity $\sigma_{\text{yy}}$ as a
function of frequency.}
\end{figure}

We numerically solve the equation of motion (\ref{yyEom}) with the
boundary conditions mentioned above and obtain $\sigma_{\text{yy}}$.
The results are plotted in Figure~\ref{Condyy} with various values
of the Gauss-Bonnet coefficient. We can clearly see from the figure
that for a fixed frequency, both the real and imaginary parts of the
conductivity $\sigma_{\text{yy}}$ decrease as the increase of the
Gauss-Bonnet coefficient. In addition, from the conductivity, one
can conclude that the ratio of gap frequency over the critical
temperature increases when the Gauss-Bonnet coefficient becomes
large.

\subsection{$\sigma_{\text{xx}}$}
The calculation of $\sigma_{\text{xx}}$ is more complicated. The
equation of motion for $A_{\text{x}}^3$ mixes it with other two
components $A_{\text{t}}^1$ and $A_{\text{t}}^2$ in
(\ref{infesimal}). The equations of motion for the three components
are
\begin{eqnarray} \label{xxEoms}
{A_{x}^3}''+(\frac1 r+\frac{f'}{f})A_{x}^3{'}+\frac{\omega ^2
A^3_{x}}{f^2}-\frac{i\omega A_{t}^2+A_{t}^1\phi}{f^2}\psi=0,\\
\label{xxEoms2}A_{t}^1{''}+\frac{3 A_{t}^1{'}}{r}+\frac{L^2 A_{x}^3
\phi \psi}{r^2
f}=0,\\
\label{xxEoms3}A_{t}^2{''}+\frac{3 A_{t}^2{'}}{r}-\frac{L^2 A_{t}^2
\psi^2}{r^2 f}-\frac{i L^2 \omega  A_{x}^3 \psi}{r^2 f}=0.
 \end{eqnarray}
Besides these three equations of motion, there are two additional
constraints:
\begin{eqnarray}\label{xxC}
-  \phi A_{t}^1{'}+ i  \omega A_{t}^2{'}-\frac{L^2 f \psi
A_{x}^3{'}}{ r^2}+
 A_{t}^1 \phi '+\frac{L^2 A_{x}^3 f \psi '}{ r^2}=0,\\
\label{xxC2}{i \omega A_{t}^1{'}}+\phi A_{t}^2{'}-{A_{t}^2 \phi
'}=0.
\end{eqnarray}
The constraint equations (\ref{xxC}) and (\ref{xxC2}) are not
independent of the equations of motion
(\ref{xxEoms}),(\ref{xxEoms2}) and (\ref{xxEoms3}). Actually, the
derivatives of the two constraint equations follow algebraically
from the equations of motion.

Again we focus on the solutions with the in-falling wave conditions
on the horizon, which determine the retarded Green's function. With
the in-falling wave condition, we get from
Eqs.~(\ref{xxEoms}),(\ref{xxEoms2}) and (\ref{xxEoms3}) that near
the horizon,
\begin{eqnarray}
A_{\text{x}}^3&=&(1-\frac{r_h}{r})^{-i\omega
L^2/(4r_h)}[1+a_{\text{x}}^{3(1)}(1-\frac{r_h}{r})+a_{\text{x}}^{3(2)}(1-\frac{r_h}{r})^2+\cdots],\\
A_{\text{t}}^1&=&(1-\frac{r_h}{r})^{-i\omega
L^2/(4r_h)}[a_{\text{t}}^{1(2)}(1-\frac{r_h}{r})^2+a_{\text{t}}^{1(3)}(1-\frac{r_h}{r})^3+\cdots],\\
A_{\text{t}}^2&=&(1-\frac{r_h}{r})^{-i\omega
L^2/(4r_h)}[a_{\text{t}}^{2(1)}(1-\frac{r_h}{r})+a_{\text{t}}^{2(2)}(1-\frac{r_h}{r})^2+\cdots].
\end{eqnarray}
where $a^{a(i)}_{\mu}$ are some constants. The boundary conditions
on the boundary of the AdS bulk can also be read from the expansion
of Eqs.~(\ref{xxEoms}), (\ref{xxEoms2}) and (\ref{xxEoms3}). They
are: \be\label{a3xbd}
 A^3_x&=&A^{3(0)}_x+\frac{A^{3(2)}_x}{r^2}+\frac{A^{3(0)}_x\om^2L_\text{eff}^4\log(\La
 r)}{2r^2}+\cdots,\\
 A^1_t&=&A^{1(0)}_t+\frac{A^{1(2)}_t}{r^2}+\cdots,\\
 A^2_t&=&A^{2(0)}_t+\frac{A^{2(2)}_t}{r^2}+\cdots.
\ee These coefficients in the expansions can be fixed using the
equations of motion and the constraint equations.

\begin{figure}
\includegraphics[width=8cm] {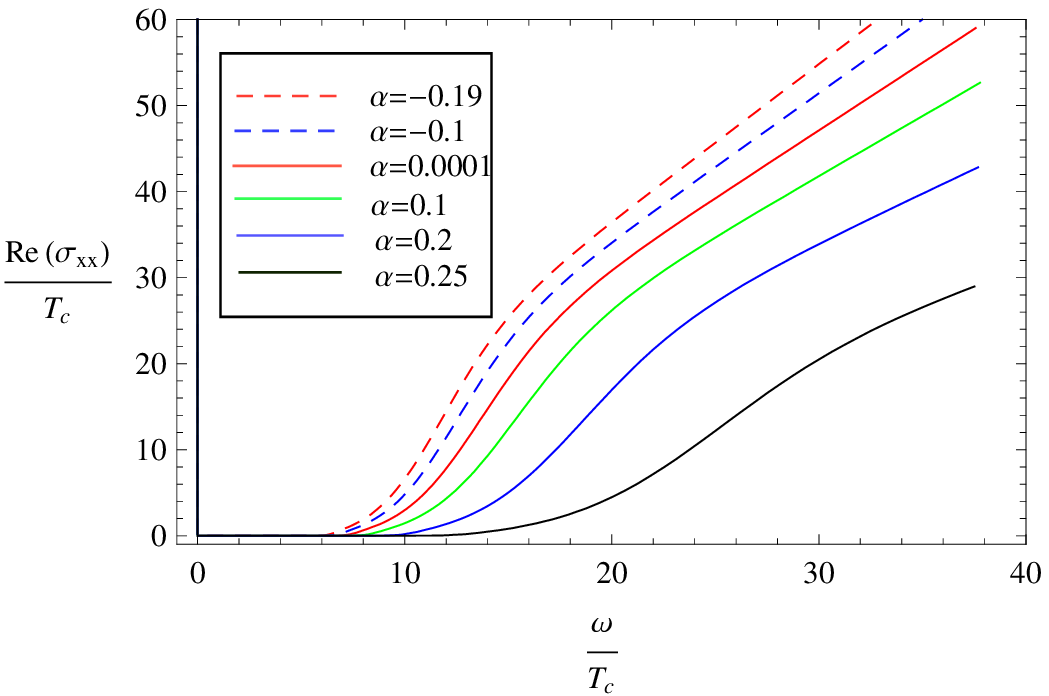}
\includegraphics[width=8cm] {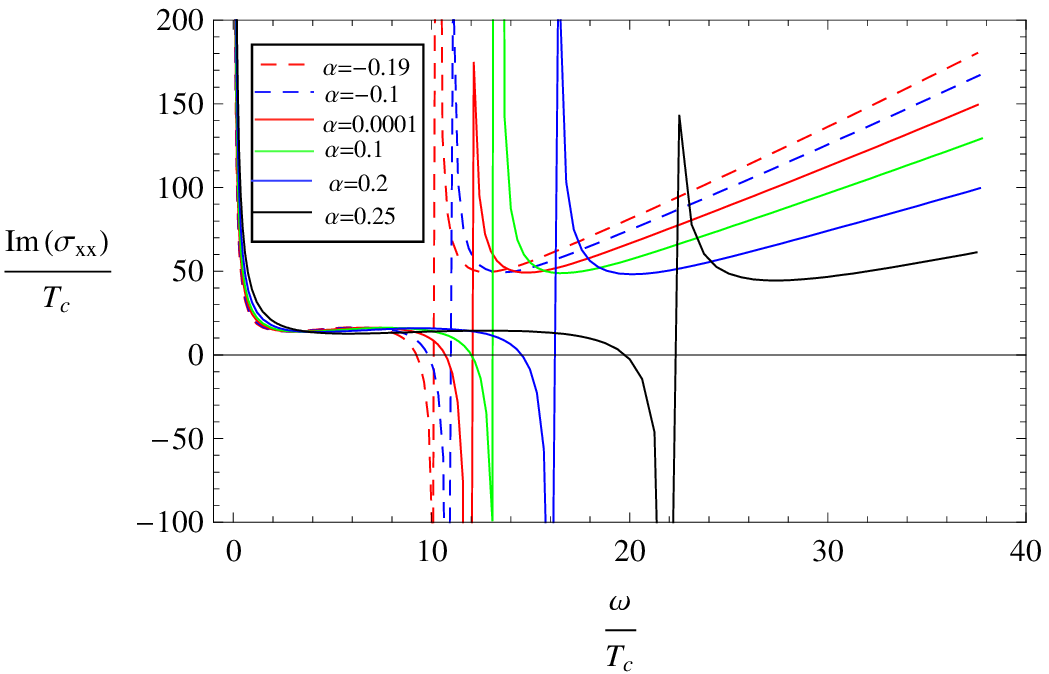}
\caption{\label{Condxx} The conductivity $\sigma_{\text{xx}}$ as a
function of frequency.}
\end{figure}

Here the calculation of conductivity is more subtle than that of
$\sigma_{\text{yy}}$. Because the definition of $A_{x}^3$ depends on
the choice of gauge, we cannot obtain the conductivity
straightforwardly by using the solution of $A_{x}^3$ only. What we
needed is a physical combination of $A_{x}^3$, $A_{\text{t}}^1$ and
$A_{\text{t}}^2$, in other words, a gauge invariant quantity. As
argued in Ref.~\cite{Gubser:2008wv}, this gauge invariant quantity
should be
\begin{equation}
\hat{A^3_x}=A^3_x+\psi \frac{i\omega L^2A^2_t + \phi
A^1_t}{\phi^2-\omega^2L^4}.
\end{equation}
This gauge invariant quantity near the boundary is
\begin{equation}
\hat{A}_x^3=A^{3(0)}_x+\frac{\hat{A}^{3(2)}_x}{r^2}+\frac{A^{3(0)}_x\om^2L_\text{eff}^4\log(\La
 r)}{2r^2},
\end{equation}
where
 \be \hat{A}^{3(2)}_x\equiv A^{3(2)}_x+\psi^{(2)}\frac{i\om
L^2A^{2(0)}_t+\mu A^{1(0)}_t}{\mu^2-\om^2L^4}.\ee With
$\hat{A}_x^3$,  we can compute the conductivity $\sigma_{\text{
xx}}$ as
\begin{equation}
\sigma_{\text{xx}}=\frac{1}{i\omega}G^R(\omega,k=0)=-\frac{2 i
\hat{A}_x^{3(2)}}{{A}_x^{3(0)}\omega L_{\text{eff}}^2 }+\frac{1}{2}
i \omega L_{\text{eff}}^2.
\end{equation}
The results are plotted in Figure~\ref{Condxx}. Comparing with
Figure~\ref{Condyy}, we can see that the conductivity
$\sigma_{\text{xx}}$ behaves quite different from
$\sigma_{\text{yy}}$. The real part of the $\sigma_{xx}$ grows much
slowly than that of $\sigma_{yy}$. The anisotropic behavior of
conductivity is just the feature of p-wave superconductors.

\subsection{More on $\si_{xx}$ and $\sigma_{\text{yy}}$}

From Figure \ref{Condyy}, we see that the real part of the
conductivity of $\sigma_{\text{yy}}$ rises quickly around some
frequency. This behavior is very similar to the case of holographic
s-wave
superconductors~\cite{Hartnoll:2008vx,Hartnoll:2008kx,Horowitz:2008bn}.
In particular, the growing behavior of the conductivity for large
frequency is due to the term $\log(\La r)$ in the expansion
(\ref{a3xbd}).  This behavior of the conductivity is the main
discrepancy between (3+1)-dimensional and (2+1)-dimensional s-wave
superconductor~\cite{Horowitz:2008bn}; in the latter case, the real
part of the conductivity approaches to a constant when $\omega \to
\infty$.  We can define $\om_g$ as the gap frequency where the
imaginary part of the conductivity Im$\sigma_{\text{yy}}$ minimizes
as in \cite{Horowitz:2008bn}, which describes excitation of
quasi-particles in pairs. The normal contribution to the DC
conductivity of $\sigma_{\text{yy}}$ is defined as
$n_n=\lim_{\omega\rightarrow0}{\text Re}(\sigma)$ which is
exponentially suppressed as $n_n\sim e^{-\Delta/T}$, where $\Delta$
is the mass of excited quasi-particles. In the BCS theory,
$\om_g=2\Delta$. However, in holographic superconductors, this
relation no longer holds in general~\cite{Horowitz:2008bn}. This
might be caused by strong coupling between quasi-particles.

From the right panel of Figure \ref{Condyy}, we read off
 \be \label{om}
 \om_{g(\al=-0.19)}&\approx& 6.6 T_c,\quad
     \om_{g(\al=-0.1)}\approx 7.0 T_c, \nno\\
 \om_{g(\al=0.0001)}&\approx& 7.7T_c,\quad
     \om_{g(\al=0.1)}\approx 8.6T_c,\nno\\
     \om_{g(\al=0.2)}&\approx& 10.5T_c,\quad
     \om_{g(\al=0.25)}\approx 14.0T_c. \ee
As in the case of  holographic s-wave superconductors, the ratio of
the gap frequency over the critical temperature  $\omega_g/T_c$
deviates from the universal value 8 and the ratio depends on the
Gauss-Bonnet coefficient. We can see that $\om_g$ increases as $\al$
grows. Because $\om_g$ can be interpreted as the energy to break a
pair of fermions, the bigger $\om_g$ is, the harder the fermion
pairs to form. Thus we can draw the conclusion again that a positive
Gauss-Bonnet term makes the condensation harder.
 Furthermore, because the boundary spacetime is four dimensional, the conductivity is of mass dimension one.
Therefore, we can read off $\Delta$ from the dimensionless quantity
Re$(\sigma)/T_c$. They are
 \be \label{delta}
 \Delta_{\alpha=-0.19}&\approx&4.62T_c,\quad
\Delta_{\alpha=-0.1}\approx4.64T_c\nno\\
 \Delta_{\alpha=0.0001}&\approx&4.69T_c,\quad
\Delta_{\alpha=0.1}\approx4.77T_c,\nno\\
\Delta_{\alpha=0.2}&\approx&4.94T_c,\quad
\Delta_{\alpha=0.25}\approx5.28T_c,\ee for different Gauss-Bonnet
coefficient. Comparing (\ref{om}) with (\ref{delta}), we can clearly
see that in this holographic model of p-wave superconductor,
$\om_g\neq2\Delta$. In particular, we notice that one has $\om_g <2
\Delta$. We can explain the difference $2\Delta -\om_g$ as the bound
energy between a pair of quasi-particles. Two exceptions are the
cases of $\alpha =0.2$ and $\alpha =0.25$, for which $\om_g
> 2\Delta$. Let us recall that the condensation behavior for the case
of $\alpha=0.25$ behaves strange (see Figure \ref{Condensation}).
Note that the causality condition for the dual field theory leads to
a constraint on the Gauss-Bonnet coefficient $-7/36 \le \alpha \le
9/100$. Thus we can conclude that strange behavior for the case of
$\alpha=0.25$ in fact demonstrates that the results for the cases
$\alpha=0.20$, and $0.25$ are not trustable since the dual field
theories are not well-defined.

\begin{figure}
\includegraphics[width=8cm] {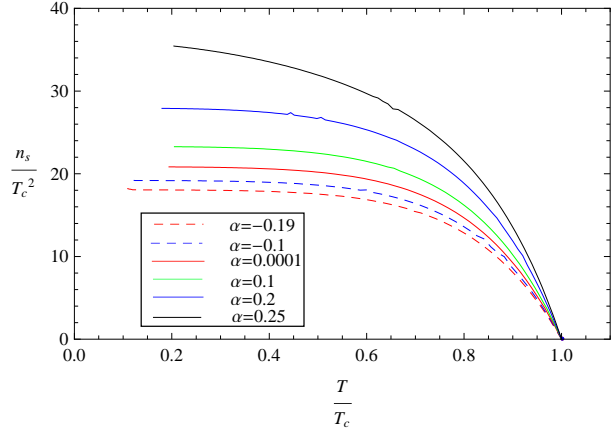}
\caption{\label{nsyy} The superfluid density of $\sigma_{yy}$ for
various $\al$.}
\end{figure}
\begin{figure}
\includegraphics[width=8cm] {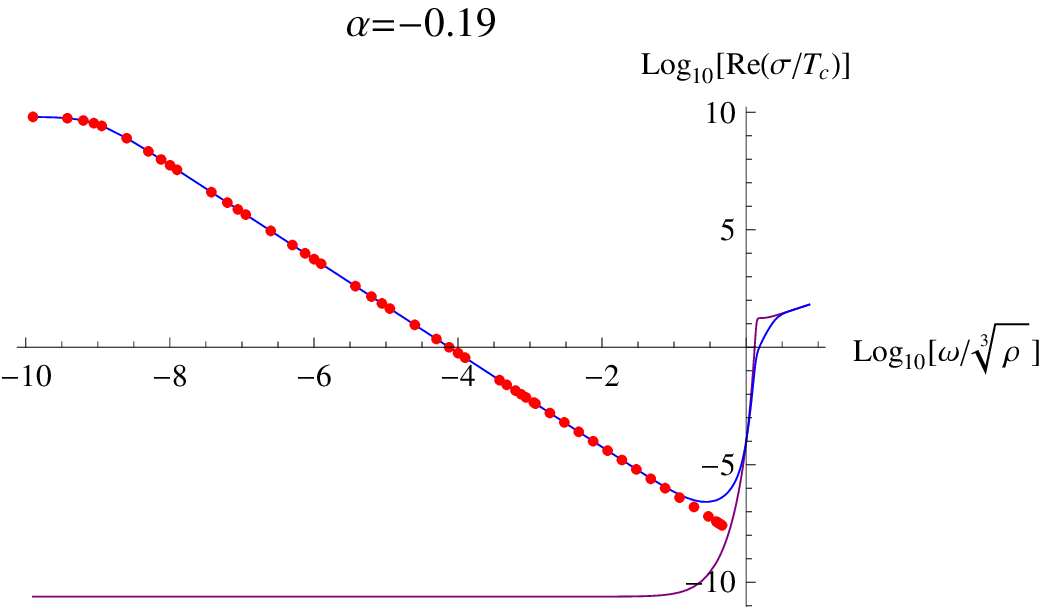}
\includegraphics[width=8cm] {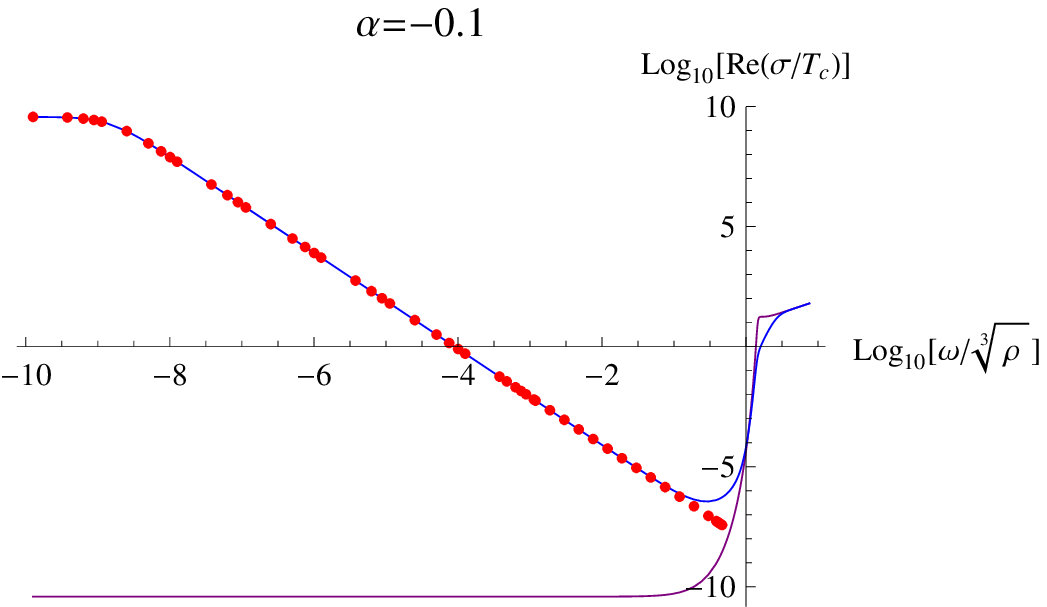}\\
\includegraphics[width=8cm] {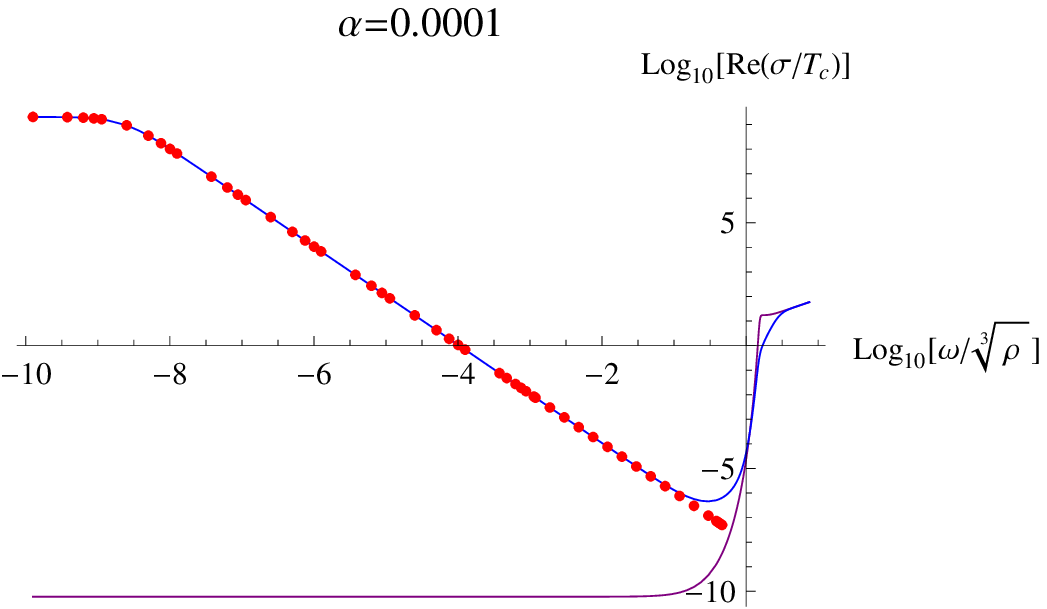}
\includegraphics[width=8cm] {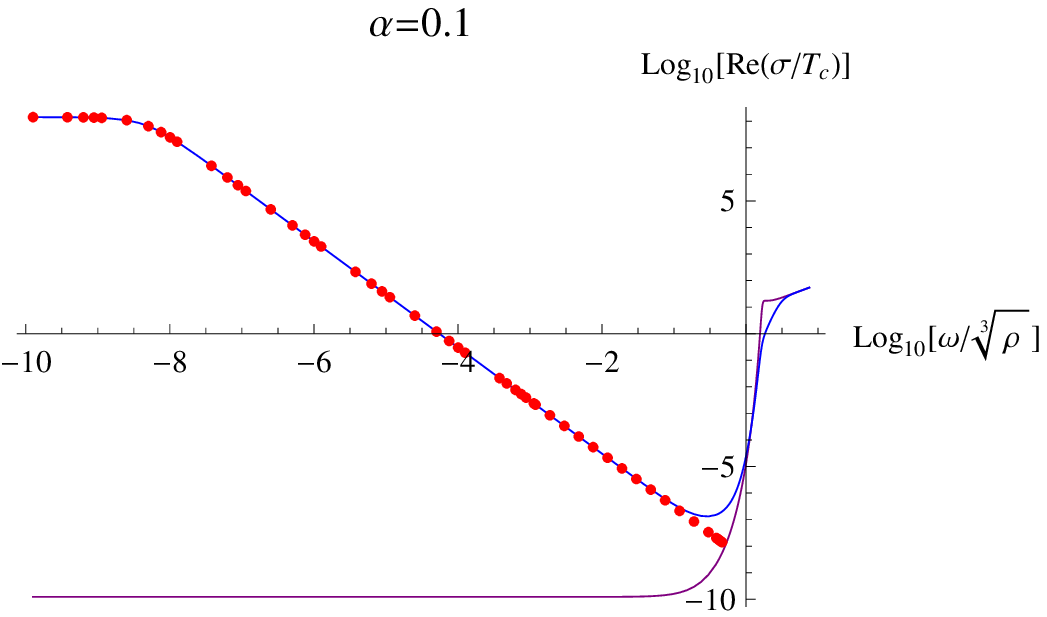}\\
\includegraphics[width=8cm] {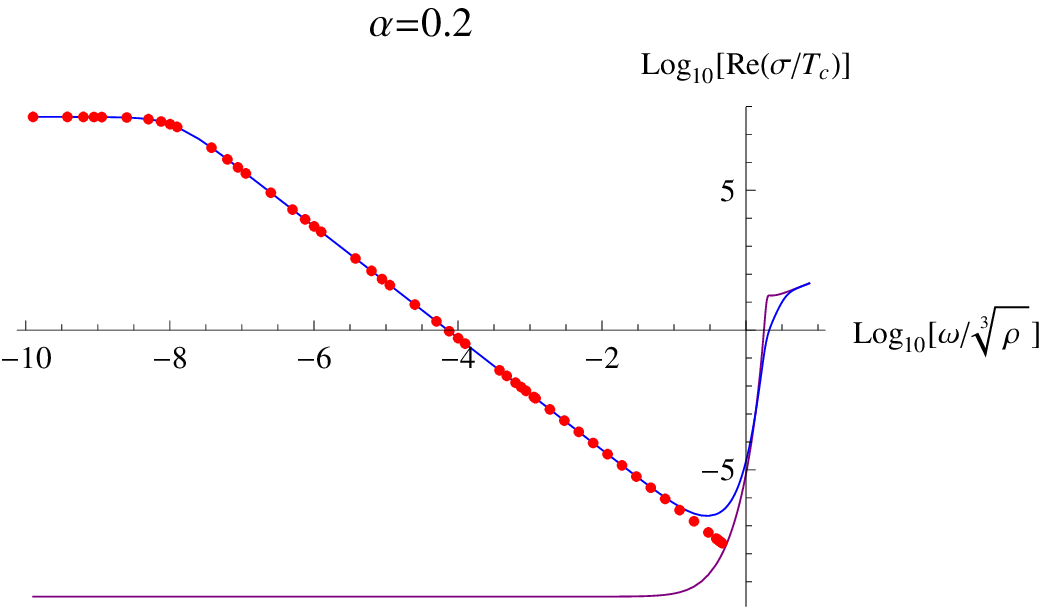}
\includegraphics[width=8cm] {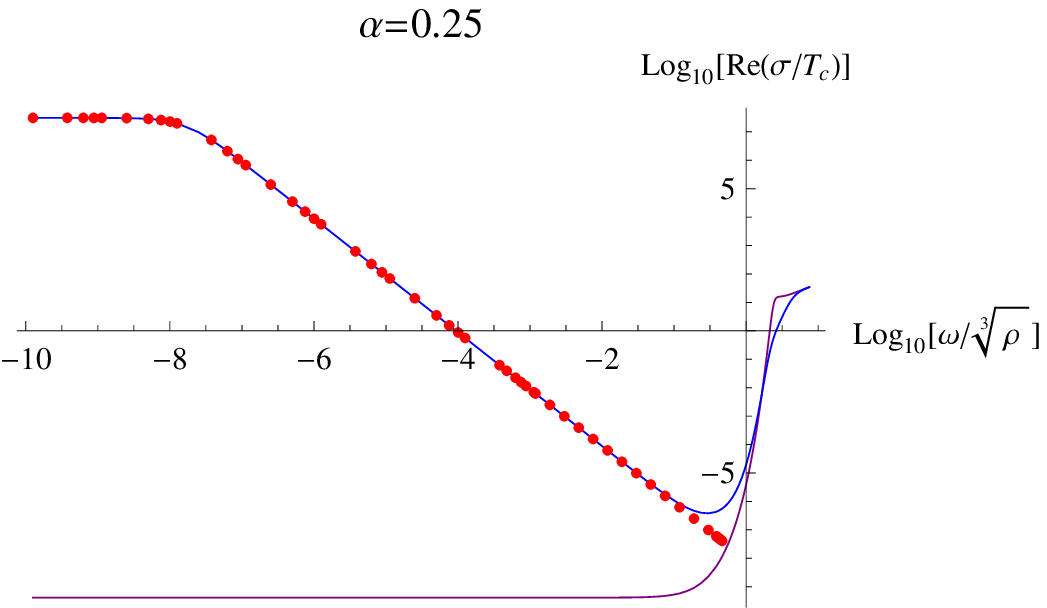}
\caption{\label{log}  Logarithmic real parts of conductivity versus
the logarithmic frequencies for various $\al$. The blue line
represents $\si_{xx}$ while the purple line for
$\sigma_{\text{yy}}$. The red points are the fitting points of
$\si_{xx}$ in the low frequency regime. }
\end{figure}

Another important order parameter of s-wave superconductors is the
superfluid density $n_s$. It can be related to the retarded Green's
function as $n_s={\text Re}[G^R(\om=k=0)]$. The behavior of the
dimensionless quantity $n_s/T_c^2$ is plotted in Figure \ref{nsyy}
for various Gauss-Bonnet coefficient $\al$. Near the critical
temperature, $T\rightarrow T_c$, the superfluid density is linearly
proportional to $(T_c-T)$ as follows:
 \be
 n_{s(\al=-0.19)}&\approx&87.54 T_c(T_c-T),\quad
     n_{s(\al=-0.1)}\approx 93.40 T_c(T_c-T)\nno\\
 n_{s(\al=0.0001)}&\approx&98.70 T_c(T_c-T),\quad
     n_{s(\al=0.1)}\approx111.14 T_c(T_c-T),\nno\\
     n_{s(\al=0.2)}&\approx&126.37 T_c(T_c-T),\quad
     n_{s(\al=0.25)}\approx146.14 T_c(T_c-T).\ee
This linear behavior is the same as in the four dimensional
case~\cite{Gubser:2008wv}. In addition we can see that the factors
of the linear relation increase as $\al$ grows. Here it shows again
that the Gauss-Bonnet term does not change the critical exponent
associated with the superfluid density.

Now let's consider $\si_{xx}$.  The behavior of the component
$\si_{xx}$ is much different from that of $\sigma_{\text{yy}}$,
which can be observed from Figure \ref{log} when the frequency is
very small. The logarithmic behavior of $\si_{xx}$ reminds us of the
Drude model which is a classical description of the electrical
conductivity in a metal:
 \be {\text Re}(\si)_{\text{Drude}}=\frac{\si_0}{1+\om^2\tau^2},\ee
where, $\si_0=ne^2\tau/m$ is the DC conductivity of $\si_{xx}$, $n,
e, m, \tau$ are respectively the electron's number density, charge,
mass and the mean free time between the ionic collision. For
$T/\sqrt[3]{\rho}\approx 0.04$, we fit the $\tau$ and $\si_0$ for
various $\al$ in low frequencies. In Figure \ref{log}, the blue
lines represent the logarithmic behavior of Re$(\si_{xx})$ while the
purple lines stand for the logarithmic behavior of
Re$\sigma_{\text{yy}}$. The red points in Figure \ref{log} are the
fitting points for Re$\si_{xx}$ in low frequencies. The values of
$\tau$ and $\si_0$ can be read from the fitting points as
 \be
\si_{0(\al=-0.19)}\approx 1.09082*10^{10} T,&& \tau_{(\al=-0.19)}
\approx 1.34592*10^8 T^{-1},\nno\\ \si_{0(\al=-0.1)}\approx
6.07532*10^9 T,&&
\tau_{(\al=-0.1)} \approx 8.55262*10^7 T^{-1} ,\nno\\
 \si_{0(\al=0.0001)}\approx 3.26748*10^9 T,&&
\tau_{(\al=0.0001)} \approx
5.47106*10^7 T^{-1} ,\nno\\
\si_{0(\al=0.1)} \approx 2.18399*10^8 T,&& \tau_{(\al=0.1)} \approx
2.72375*10^7 T^{-1},\nno\\
 \si_{0(\al=0.2)} \approx 6.11933*10^7 T, &&
\tau_{(\al=0.2)} \approx 1.13346*10^7 T^{-1} ,\nno\\
\si_{0(\al=0.25)} \approx 4.22257*10^7 T,&& \tau_{(\al=0.25)}
\approx 7.43723*10^6 T^{-1}.\ee
 We see that in general, both the DC conductivity and the mean free time $\tau$
 decrease as $\al$ grows.

\section{Conclusions}
\label{sect:conclu}

In this paper we studied the holographic p-wave superconductors in a
five-dimensional Einstein-Gauss-Bonnet gravity theory with an SU(2)
Yang-Mills gauge field. We treated the SU(2) Yang-Mills field as a
probe field, which means the back reaction of the Yang-Mills field
on the background is not taken into account.  A component of the
vector field will condense when the temperature of the Gauss-Bonnet
black hole is below a critical value. This condensation is
interpreted as a p-wave superconducting phase transition on the
boundary field theory. We found that when the Gauss-Bonnet coupling
increases, the value of the condensation becomes bigger and the
critical temperature decreases, as in the case of holographic s-wave
superconductors. This means a positive Gauss-Bonnet term  makes the
condensation harder. This phenomena is also  observed from the gap
frequency $\omega_g$ which increases as the Gauss-Bonnet coupling
grows.

The electric conductivity of the p-wave superconductor is quite
different from the one for s-wave superconductors. The conductivity
perpendicular to the direction of the condensation behaves like a
s-wave one, while the conductivity parallel to the direction of the
condensation behaves much different. In the low frequency regime,
this conductivity can be well explained by the Drude model. As for
the effect of the Gauss-Bonnet term,  both the DC conductivity and
the mean free time decrease as $\al$ increases. In the low frequency
regime, we obtained the mass of excited quasi-particles $\Delta$ for
$\sigma_{\text{yy}}$. We found that in this holographic model, the
usual relation $\om_g=2\Delta$ in BCS theory  does not hold, which
demonstrates the strong coupling between excited quasi-particles.
For the conductivity $\si_{xx}$, fitting the data, we obtained the
DC conductivity and mean free time in the low frequency regime.

In particular, we observed that the condensation for the case of
$\alpha=0.25$ and the relation between the gap frequency and the
mass of quasi-particles for the cases of $\alpha=0.20$ and $0.25$
behaves strange. This strange behavior is consistent with the
causality bound on the Gauss-Bonnet coefficient from the dual field
theory~\cite{Brigante:2007nu,Brigante:2008gz}. The latter imposes a
constrain on the coefficient: $-7/36 \leq \alpha \leq 9/100$. This
implies that our results for the cases of $\alpha > 9/100$ are not
trustable.

Note that here the superconducting phase transition is still second
order. Namely the Gauss-Bonnet term does not change the order of the
phase transition and some critical exponents. They still take the
mean-field theory values.  On the other hand, the back reaction of
the SU(2) Yang-Mills field will change the phase transition from
second order to first order when the ratio of the gravitation
constant to the Yang-Mills coupling reaches a critical value
\cite{Ammon:2008fc}. It would be interesting to see how the
Gauss-Bonnet term affects the phase transition when the back
reaction is included.

\section*{Acknowledgements}
RGC  thanks the organizers and participants for various discussions
during the workshop on ¡°Dark Energy and Fundamental Theory¡± held
at Xidi, Anhui, China, May 28-June 6, 2010, supported by the Special
Fund for Theoretical Physics from the National Natural Science
Foundation of China with grant No: 10947203. ZYN and HQZ would like
to thank M.M. Roberts and B. Hu for helpful discussions. ZYN would
like to thank B.N. Lu and Y.W. Sun for useful discussions. This work
was partially supported by the National Natural Science Foundation
of China (No. 10821504 and No. 10975168).

\end{document}